\documentclass[aps,prb,twocolumn]{revtex4-2}

\usepackage{amsmath}
\usepackage{amssymb}
\usepackage{graphicx}
\usepackage{dcolumn}
\usepackage{bm}

\newcommand{\dd}{\mathrm{d}}

\newcommand{\mbf}[1]{\mathbf{#1}}

\newcommand{\mr}[1]{\mathrm{#1}}
\newcommand{\pd}{\partial}

\begin{document}
\title{Scaling and data collapse of two-dimensional random singlet state in magnetic field}

\author{Chen Peng}
\affiliation{Kavli Institute for Theoretical Sciences and CAS Center for Excellence in Topological Quantum Computation, University of Chinese Academy of Sciences, Beijing 100190, China}

\author{Long Zhang}
\email{longzhang@ucas.ac.cn}
\affiliation{Kavli Institute for Theoretical Sciences and CAS Center for Excellence in Topological Quantum Computation, University of Chinese Academy of Sciences, Beijing 100190, China}

\date{\today}

\begin{abstract}
Quenched randomness strongly affects properties of magnetic materials. Two-dimensional (2D) random singlet (RS) states emerge in random $J$-$Q$ model by destroying valence bond solid order with spatial randomness. We examine the 2D RS state in magnetic field with quantum Monte Carlo simulations. The magnetization and susceptibilities show power-law scaling with magnetic field at low temperature. Moreover, they show one-parameter scaling behavior with $B/T$, and the scaling functions are remarkably consistent with a phenomenological model of random spin pairs with a singular distribution of interactions. These universal scaling functions can be used to diagnose 2D RS states in experiments.
\end{abstract}

\maketitle

{\bf Introduction.} Quenched randomness is inevitably introduced by impurities and defects in quantum magnetic materials, and can strongly influence their physical properties. In quasi-one-dimensional (quasi-1D) antiferromagnetic (AF) materials, e.g., the organic charge-transfer salts tetracyanoquinodimethanide (TCNQ) compounds \cite{Bulaevskii1972, Shchegolev1972, Azevedo1977}, the specific heat coefficient and the uniform magnetic susceptibility are significantly enhanced in power law at low temperature, $C/T\sim T^{-\alpha_{c}}$ and $\chi\sim T^{-\alpha_{s}}$ with $0<\alpha_{c,s}<1$, implying a divergent low-energy density of states. This inspired the phenomenological model with a singular distribution of density of states \cite{Bulaevskii1972} and random exchange interactions \cite{Theodorou1976, Theodorou1977}, $P(J)\sim J^{-\alpha}$, although without prior justification.

In the seminal work by Ma and Dasgupta \cite{Ma1979, Dasgupta1980}, it was shown that in the spin-1/2 random Heisenberg chain, such singular distribution of low-energy effective exchange interactions emerges universally with the real-space renormalization group (RSRG) transformations. The ground state is captured by the random singlet (RS) state \cite{Bhatt1981a, Bhatt1982, Fisher1994} composed of a bunch of spin singlet pairs, which are decimated from the spin chain during the RSRG transformations. The RSRG approach was later put on firm ground \cite{Fisher1994}, because the distribution of the logarithm of effective exchange interactions becomes broader and broader with the RG flow, and approaches the infinite-randomness fixed point (IRFP). The IRFP is characterized by highly anisotropic dynamical scaling behavior, in which the characteristic energy and length scales are related by $\ln(1/\epsilon)\sim \sqrt{l}$, implying an infinite dynamical exponent $z$.

In the past two decades, the experimental pursuit of quantum spin liquid (QSL) materials \cite{Lee2008, Balents2010, Zhou2017} found similar power-law scaling of thermodynamic quantities at low temperature in several quasi-2D QSL candidate materials, e.g., the rare-earth compound YbMgGaO$_{4}$ \cite{Li2015}. While this might be attributed to the postulated gapless QSL states with spinon Fermi surfaces and emergent gauge fluctuations \cite{Shen2016b, Li2017e}, quenched randomness together with geometric frustration can destroy magnetic orders and lead to a divergent low-energy density of states as well \cite{Zhu2017d, Kimchi2018a, Kimchi2018, Parker2018a}. The absence of coherent spinon thermal transport at low temperature \cite{Xu2016d} further strengthens the argument for quenched randomness, although a general consensus has not been reached \cite{Li2019d, Rao2021}. Therefore, disordered ground states induced by quenched randomness in 2D magnetic materials deserve further study. However, numerical studies of 2D random spin models with geometric frustration by exact diagonalization or density-matrix renormalization group methods are restricted to small lattice size or quasi-1D lattice geometry \cite{Shimokawa2015, Zhu2017d, Uematsu2017a, Uematsu2018, Uematsu2019, Kawamura2019, Wu2019a, Wu2021, Ren2023}.

In a recent work \cite{Liu2018b, Liu2020}, 2D RS ground states are found in the random $J$-$Q$ model, which is tractable with sign-problem-free quantum Monte Carlo (QMC) simulations. The uniform $J$-$Q$ model defined in Eq. (\ref{eq:JQ}) was introduced \cite{Sandvik2007, Lou2009a} to study the postulated deconfined quantum critical point \cite{Senthil2004b, Senthil2004a} bridging the AF order in the small $Q/J$ regime to the valence bond solid (VBS) order in the large $Q/J$ regime. The quenched spatial randomness linearly couples to the VBS order parameter, thus immediately destroys the long-range VBS order \cite{Imry1975, Binder1983, Nattermann1988} and gives way to the 2D RS state, which was proposed in Ref. \cite{Kimchi2018a} and confirmed numerically in Refs. \cite{Liu2018b, Liu2020}. The specific heat coefficient and the susceptibility show power-law dependence on temperature, $C/T\sim T^{-\alpha_{c}}$ and $\chi\sim T^{-\alpha_{s}}$ with $0<\alpha_{c,s}<1$. However, the exponents $\alpha_{s,c}$ continuously vary within the RS phase, which was attributed to a finite and nonuniversal dynamical exponent $z>2$ \cite{Liu2018b, Liu2020} in sharp contrast to the 1D RS state.

In this work, we examine the 2D RS state in external magnetic field with unbiased large-scale QMC simulations and scaling analysis. The magnetic field polarizes spins and suppresses their quantum and thermal fluctuations, thus serves as an efficient tuning knob of magnetic materials. In a phenomenological model of RS states \cite{Bhatt1981a, Bhatt1982, Kimchi2018a} composed of an ensemble of random spin pairs with a singular distribution of effective exchange interactions, $P(\tilde{J})\sim \tilde{J}^{-\alpha}$, spin pairs with $\tilde{J}<B$ are polarized at the ground state, thus the magnetization $m_{z}\propto B^{1-\alpha}$, while the susceptibility $\chi\propto B^{-\alpha}$. At finite temperature, the rescaled quantities $\tilde{m}_{z}(B,T)=T^{\alpha-1}m_{z}(B,T)$ and $\tilde{\chi}(B,T)=T^{\alpha}\chi(B,T)$ collapse onto universal scaling functions of the ratio $B/T$ given in Eqs. (\ref{eq:Mtilde})--(\ref{eq:chittilde}). The power-law scaling and data collapse behavior is also consistent with the quantum critical scenario, which assumes that the 2D RS state is captured by a scale-invariant fixed point.

We first confirm that the magnetization and the susceptibilities of the 2D RS state show power-law dependence on magnetic field in the low-temperature limit. Moreover, we show that these thermodynamic quantities show one-parameter scaling and data collapse behavior with the ratio $B/T$, and the universal scaling functions are remarkably consistent with those derived from the phenomenological model of random spin pairs. This is beyond the power-law scaling form given in Refs. \cite{Kimchi2018a, Kimchi2018}, which is valid only in the limits $B/T\gg 1$ and $B/T\ll 1$. Therefore, it provides a benchmark for future experimental and theoretical study of possible 2D RS states in quantum magnetic materials.

This paper is organized as follows. We first introduce the random $J$-$Q$ model and the physical quantities evaluated with QMC simulations, and the quantum critical scaling scenario and the phenomenological model. We then present the numerical results and scaling analysis, and conclude with a brief summary and discussions.

\begin{figure}[tb]
\centering
\includegraphics[width=\columnwidth]{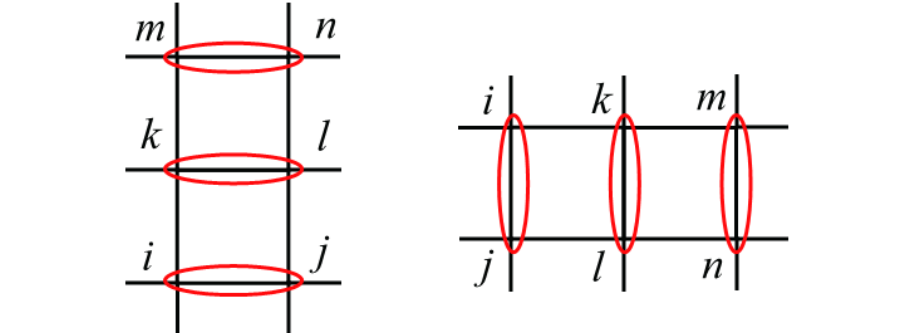}
\caption{Illustration of the six-spin interaction $Q$-term in the $J$-$Q$ model on a square lattice defined in Eq. (\ref{eq:JQ}), in which the projection operators $P_{ij}$'s act on the encircled spin pairs.}
\label{fig:lat}
\end{figure}

{\bf Model and method.} The $J$-$Q$ model on a square lattice is defined by \cite{Sandvik2007, Lou2009a}
\begin{equation}
H_{0}=-J\sum_{\langle ij\rangle}P_{ij}-Q\sum_{[ijklmn]}P_{ij}P_{kl}P_{mn},
\label{eq:JQ}
\end{equation}
in which $P_{ij}=1/4-\mbf{S}_{i}\cdot\mbf{S}_{j}$ is the projection operator into the singlet state of the spin pair. The first term is the AF exchange interactions of nearest-neighbor spins. The second term is the six-spin interactions over all columnar aligned spin pair triples illustrated in Fig. \ref{fig:lat}. The $Q$-term favors the columnar VBS order, which spontaneously breaks the lattice translation and rotation symmetry. The quantum phase transition from the AF order to the VBS order in the $J$-$Q$ model has been extensively studied with QMC simulations \cite{Sandvik2007, Lou2009a, Sandvik2010b, Shao2016a}.

Quenched spatial randomness is introduced by taking the interaction strengths $J$ and/or $Q$ at different positions as independent random variables. Following Refs. \cite{Liu2018b, Liu2020}, we choose a uniform $J=1$ as the unity of energy scale, and draw $Q$-terms from the following bimodal distribution,
\begin{equation}
P(Q)=
\begin{cases}
1/2,& Q=2\Lambda,\\
1/2,& Q=0.
\end{cases}
\end{equation}
Increasing $\Lambda$ enhances both the local VBS order and the spatial randomness, and induces a quantum phase transition from the AF order to the RS phase at $\Lambda_{c}=1.24(13)$ \cite{Liu2018b}. A uniform magnetic field is introduced into the Hamiltonian by the Zeeman term,
\begin{equation}
H_{B}=-B\sum_{i}S_{i}^{z}.
\end{equation}

The uniform magnetization $m_{z}$ is defined by
\begin{equation}
m_{z}=\frac{1}{N}\sum_{i}\langle S_{i}^{z}\rangle,
\end{equation}
in which the summation is taken over all $N=L^{2}$ lattice sites, and $\langle\cdot\rangle$ denotes the thermodynamic average at temperature $T$. The longitudinal and the transverse susceptibilities are inequivalent in magnetic field. The longitudinal susceptibility is given by
\begin{equation}
\chi_{L}=\frac{1}{NT}\bigg(\Big\langle\Big(\sum_{i}S_{i}^{z}\Big)^{2}\Big\rangle-\Big\langle\sum_{i}S_{i}^{z}\Big\rangle^{2}\bigg),
\end{equation}
while the transverse susceptibility is
\begin{equation}
\chi_{T}=\frac{1}{4N}\int_{0}^{1/T}\dd \tau\,\sum_{ij}\Big\langle S_{i}^{+}(\tau)S_{j}^{-}(0)+\mr{H.c.}\Big\rangle,
\end{equation}
in which $S^{\pm}(\tau)=e^{-\tau H}S^{\pm}e^{\tau H}$ is the imaginary-time evolved spin ladder operators.

The stochastic series expansion (SSE) with the directed-loop update algorithm \cite{Sandvik1992a, Sandvik1999, Syljuasen2002} is adopted to evaluate these physical quantities on square lattice with $N=L^{2}$ sites and periodic boundary condition. These quantities are evaluated with the improved estimators \cite{Sandvik2010}. All results are averaged over $N_{s}=128$ independent samples of spatial randomness.

{\bf Theoretical scenarios.} Numerical results will be analyzed based on two complementary scenarios, the quantum critical scaling and the phenomenological model of random spin pairs.

Suppose that the 2D RS state is a quantum critical state corresponding to a scale-invariant RG fixed point, and temperature and magnetic field are relevant scaling variables, then under a scale transformation, the singular part of a physical quantity $X(B,T)$ satisfies \cite{Cardy1996scaling}
\begin{equation}
X(B,T)=b^{y_{X}}X(b^{y_{B}}B, b^{y_{T}}T),
\end{equation}
in which $b$ is the scale transformation parameter, and $y_{X}$, $y_{B}>0$ and $y_{T}>0$ are the scaling dimensions of $X$, $B$ and $T$, respectively. Eliminating the parameter $b$, we find the following one-parameter scaling form,
\begin{equation}
\begin{split}
X(B,T) &=T^{-y_{X}/y_{T}}\tilde{X}_{1}(B/T^{y_{B}/y_{T}}) \\
&=B^{-y_{X}/y_{B}}\tilde{X}_{2}(T/B^{y_{T}/y_{B}}),
\end{split}
\label{eq:y}
\end{equation}
in which $\tilde{X}_{1}$ and $\tilde{X}_{2}$ are nonsingular scaling functions. In particular, we have
\begin{equation}
X(T)\propto T^{-y_{X}/y_{T}}
\label{eq:Tscaling}
\end{equation}
in the absence of magnetic field, and
\begin{equation}
X(B)\propto B^{-y_{X}/y_{B}}
\label{eq:Bscaling}
\end{equation}
in the low-temperature limit. The specific heat and the magnetic susceptibility of 2D RS states in the random $J$-$Q$ model indeed show power-law dependence on temperature \cite{Liu2018b, Liu2020}.

On the other hand, the ground state and low-energy excitations of the RS phase can be described by the phenomenological model \cite{Bhatt1981a, Bhatt1982, Kimchi2018a} of an ensemble of random spin pairs,
\begin{equation}
H_{\mr{eff}}=\sum_{l}\tilde{J}_{l}\mbf{S}_{l_{1}}\cdot\mbf{S}_{l_{2}},
\end{equation}
in which the effective exchange interactions satisfy the following singular distribution up to some energy cutoff $\Omega>0$,
\begin{equation}
P(\tilde{J})=
\begin{cases}
\frac{1-\alpha}{\Omega^{1-\alpha}}\tilde{J}^{-\alpha},& 0\leq \tilde{J}\leq \Omega, \\
0,& \mr{otherwise},
\end{cases}
\end{equation}
with the exponent $0<\alpha<1$. In a uniform magnetic field, the partition function is given by
\begin{equation}
Z(B,T)=\prod_{l}\Big(1+e^{-\tilde{J}_{l}/T}\big(1+2\cosh(B/T)\big)\Big).
\end{equation}
The magnetization is given by
\begin{equation}
\begin{split}
m_{z}(B,T) &=\frac{1}{N}\sum_{l}\frac{2e^{-\tilde{J}_{l}/T}\sinh(B/T)}{1+e^{-\tilde{J}_{l}/T}\big(1+2\cosh(B/T)\big)} \\
&= \frac{1}{2}\int_{0}^{\Omega}\dd \tilde{J}\,P(\tilde{J})\frac{2e^{-\tilde{J}/T}\sinh(B/T)}{1+e^{-\tilde{J}/T}\big(1+2\cosh(B/T)\big)}.
\end{split}
\end{equation}
Defining $x=\tilde{J}/T$ as the new variable of integration and sending its upper bound $\Omega/T\gg 1$ to infinity, we find
\begin{equation}
\begin{split}
m_{z}(B,T) &= -cT^{1-\alpha}\frac{\sinh(B/T)\mr{Li}_{1-\alpha}(-1-2\cosh(B/T))}{1+2\cosh(B/T)} \\
&\equiv T^{1-\alpha}\tilde{m}_{z}(B/T),
\end{split}
\label{eq:Mtilde}
\end{equation}
in which $c=\Omega^{\alpha-2}\Gamma(2-\alpha)$ is a proportional constant, and $\mr{Li}_{n}(z)$ is the polylogarithm function. Using its asymptotic expansion \footnote{{https://en.wikipedia.org/wiki/Polylogarithm}}, we find the following asymptotic behavior: $m_{z}(B,T)\sim BT^{-\alpha}$ for $B\ll T\ll J,\Omega$, and $m_{z}(B,T)\sim B^{1-\alpha}$ for $T\ll B\ll J,\Omega$. This phenomenological model is consistent with the quantum critical scaling scenario if the scaling dimensions satisfy $y_{B}=y_{T}$, and yields an explicit expression of the universal scaling function $\tilde{m}_{z}(B/T)$. The longitudinal susceptibility $\chi_{L}(B,T)=\pd m_{z}(B,T)/\pd B$, thus we have
\begin{equation}
\chi_{L}(B,T)= T^{-\alpha}\tilde{m}_{z}'(B/T).
\label{eq:chitilde}
\end{equation}
The transverse susceptibility satisfies $\chi_{T}(B,T)=m_{z}(B,T)/B$, thus
\begin{equation}
\chi_{T}(B,T)= T^{-\alpha}(B/T)^{-1}\tilde{m}_{z}(B/T).
\label{eq:chittilde}
\end{equation}
Therefore, we find $\chi_{L,T}(B,T)\sim T^{-\alpha}$ for $B\ll T\ll J,\Omega$, and $\chi_{L,T}(B,T)\sim B^{-\alpha}$ for $T\ll B\ll J,\Omega$. The scaling functions Eqs. (\ref{eq:Mtilde})--(\ref{eq:chittilde}) will be compared with the following numerical results.

\begin{figure}[tb]
\centering
\includegraphics[width=\columnwidth]{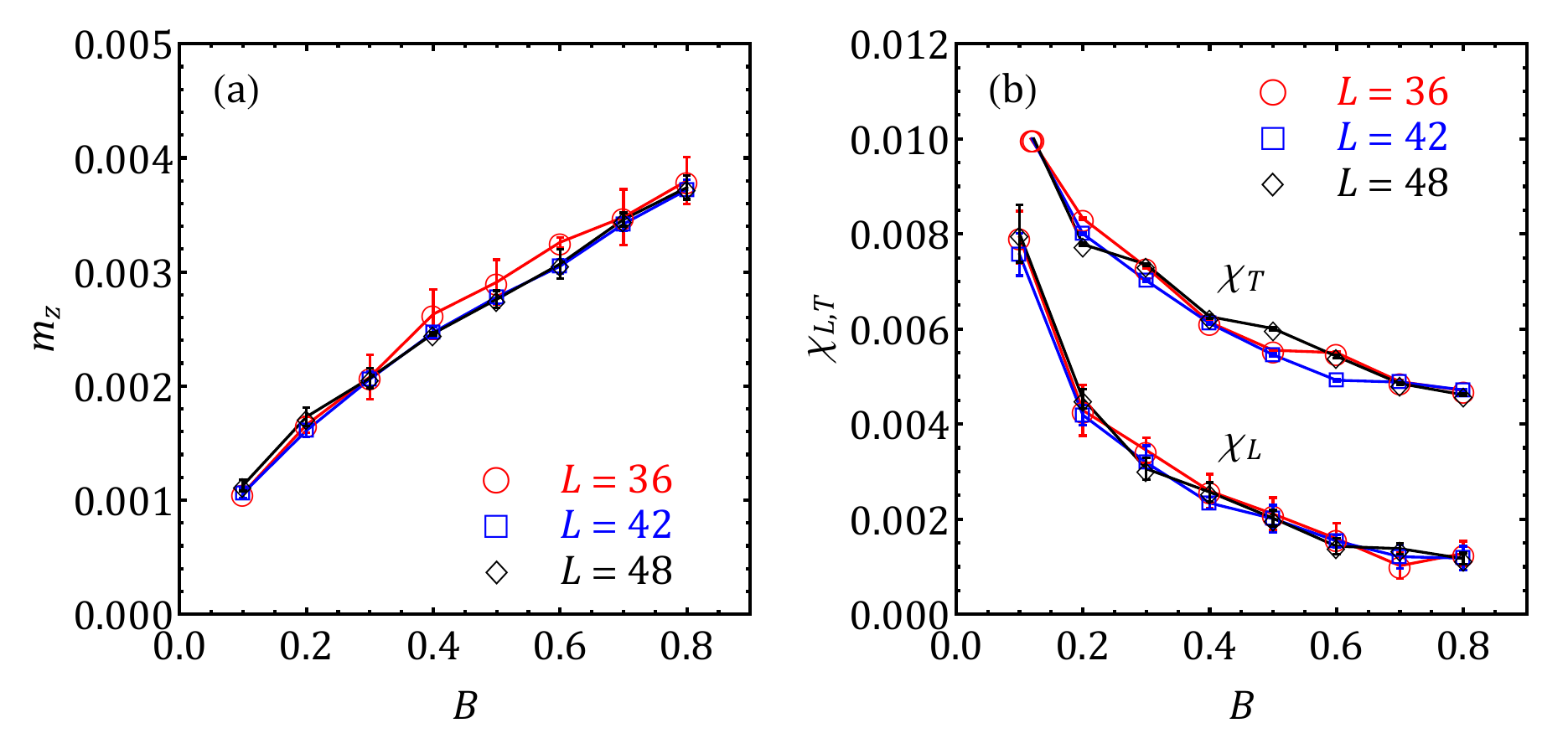}
\caption{(a) Magnetization $m_{z}$ and (b) susceptibilities $\chi_{L,T}$ versus magnetic field $B$ at temperature $T=0.04$ evaluated with different lattice sizes, $L=36$, $42$ and $48$.}
\label{fig:finite-size-effect}
\end{figure}

{\bf Numerical results and scaling analysis.} We focus on the random $J$-$Q$ model at $\Lambda=4$, which is deep in the RS phase \cite{Liu2018b}. The magnetization and the susceptibilities versus magnetic field at $T=0.04$ are plotted in Fig. \ref{fig:finite-size-effect}. We find that the results evaluated with different lattice sizes $L=36$, $42$ and $48$ overlap with each other within error bars. This indicates that the finite-size effect of these thermodynamic quantities is relatively weak in the 2D RS state, and the thermodynamic limit is essentially reached. Therefore, we take $L=42$ in the rest of this work.

\begin{figure*}[tb]
\centering
\includegraphics[width=\textwidth]{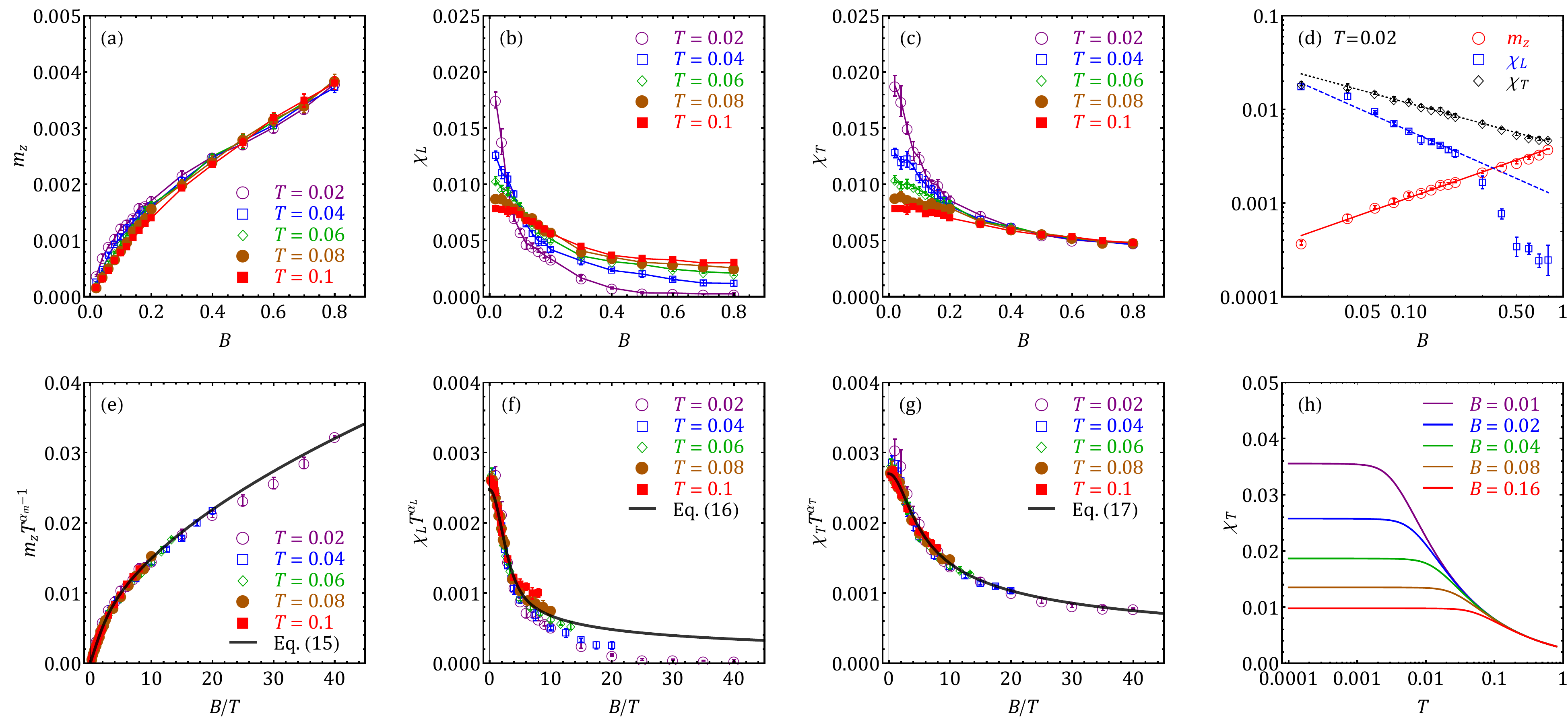}
\caption{(a--c) Magnetization $m_{z}$ and susceptibilities $\chi_{L,T}$ versus magnetic field evaluated in the temperature range $0.02\leq T\leq 0.1$. (d) Power-law fitting of magnetization $m_{z}$ and susceptibilities $\chi_{L,T}$ at the lowest temperature $T=0.02$. (e--g) Data collapse behavior of the rescaled thermodynamic quantities versus the ratio $B/T$. Thick black curves are the phenomenological scaling functions in Eqs. (\ref{eq:Mtilde})--(\ref{eq:chittilde}), in which the exponent $\alpha$ is obtained from the data collapse fitting procedure, thus there are no more free parameters in plotting these curves except for the overall proportional constant $c$. All data are evaluated with lattice size $L=42$. (h) Transverse susceptibility $\chi_{T}$ of the phenomenological model versus temperature. At nonzero magnetic field, $\chi_{T}(T)$ reaches a plateau at low temperature, and its height scales as $B^{-\alpha}$.}
\label{fig:data}
\end{figure*}

The magnetization and the susceptibilities are evaluated in the range of temperature and magnetic field $0.02\leq T\leq 0.1$ and $0.02\leq B\leq 0.8$. The results are plotted in Fig. \ref{fig:data} (a--c). According to the quantum critical scaling scenario, these quantities are expected to show power-law dependence on magnetic field in the low-temperature limit, $T\ll B\ll J,\Lambda$, thus we fit the data at $T=0.02$ with the power law, $m_{z}(B)\propto B^{1-\alpha_{m}}$ and $\chi_{L,T}(B)\propto B^{-\alpha_{L,T}}$, which yields the exponents $\alpha_{m}=0.42(2)$, $\alpha_{L}=0.7(2)$ and $\alpha_{T}=0.45(2)$. The results are plotted in Fig. \ref{fig:data} (d). Here, the fitting of $\chi_{L}$ is restricted to the range $0.02\leq B\leq 0.3$ because of the obvious deviation from the power law at higher magnetic field. The cusp in $\chi_{L}$, i.e., a discontinuity in its slope, may reflect a quantum phase transition induced by the magnetic field, which is left for future study. Error bars are estimated by combining the statistical uncertainty of the fitted exponents and their variation if the data at $B=0.02$ are excluded in the fitting procedure. We also consider a different fitting procedure by including a subleading correction term to the power-law scaling,
\begin{gather}
m_{z}(B)=a B^{1-\alpha_{m}}+a'B^{2-\alpha_{m}}, \\
\chi_{L,T}(B)=b_{L,T}B^{-\alpha_{L,T}}+b'_{L,T}B^{1-\alpha_{L,T}},
\end{gather}
in which the $a'$ and the $b'_{L,T}$ terms capture the subleading corrections from irrelevant operators near the presumable RS fixed point. We find $\alpha_{m}=0.42(2)$, $\alpha_{L}=0.58(7)$, and $\alpha_{T}=0.40(4)$. The exponents from different fitting procedures are consistent with each other, and they respect the scaling law $\alpha_{m}=\alpha_{L}=\alpha_{T}$ within one or two standard errors. Moreover, these exponents are close to that obtained in the finite-temperature scaling of the susceptibility, $\chi(T)\propto T^{-\gamma}$ with $\gamma=0.60(10)$ \cite{Liu2018b}. This suggests that the scaling dimensions $y_{B}\simeq y_{T}$ [see Eqs. (\ref{eq:Tscaling})--(\ref{eq:Bscaling})], and implies the possible one-parameter scaling and data collapse behavior with the scaling variable $B/T$, which we will examine below.

\begin{ruledtabular}
\begin{table}[tb]
\caption{Details of the data collapse analysis of magnetization and susceptibilities. Error bars in the exponent $\alpha$ reflect the statistical uncertainty of the data collapse fitting procedure. $\chi^{2}/\mr{d.o.f.}$ listed in the last column are close to unity, indicating quite reasonable quality of fitting.}
\label{tab:collapse}
\begin{tabular}{ccccc}
Quantity	& $B$ range				& $k_{\mr{max}}$	& $\alpha$	& $\chi^{2}/\mr{d.o.f.}$	\\
\hline
$m_{z}$		& $0.02\leq B\leq 0.8$	& 5					& 0.452(9)	& 1.73						\\
$\chi_{L}$	& $0.02\leq B\leq 0.3$	& 3					& 0.478(15)	& 1.50						\\
$\chi_{T}$	& $0.02\leq B\leq 0.8$	& 5					& 0.465(9)	& 1.53
\end{tabular}
\end{table}
\end{ruledtabular}

We make the following one-parameter scaling ansatz for the magnetization,
\begin{equation}
m_{z}(B,T)=T^{1-\alpha_{m}}\sum_{k=0}^{k_{\mr{max}}}c_{k}(B/T)^{k},
\end{equation}
in which the nonsingular scaling function is expanded into power series truncated at the $k_{\mr{max}}$-th order. The exponent $\alpha_{m}$ and the coefficients $c_{k}$'s are fitting parameters. We make the scaling ansatz for magnetic susceptibilities likewise,
\begin{equation}
\chi_{L,T}(B,T)=T^{-\alpha_{L,T}}\sum_{k=0}^{k_{\mr{max}}}c_{k}(B/T)^{k}.
\end{equation}
This fitting procedure is widely used for one-parameter scaling analysis in numerical simulations \cite{Sandvik2010, Deng2005} without assuming the specific form of scaling functions in advance. The truncation order $k_{\mr{max}}$ is chosen such that the reduced chi-squared statistics $\chi^2/\mr{d.o.f.}$ is close to 1 \cite{Sandvik2010}. The fitting results are listed in Table \ref{tab:collapse}. The exponents $\alpha_{m}=0.452(9)$, $\alpha_{L}=0.478(15)$ and $\alpha_{T}=0.465(9)$ are equal to each other within error bars. The rescaled quantities are plotted in Fig. \ref{fig:data} (e--g), and exhibit data collapse behavior with the scaling variable $B/T$ as expected. Moreover, we find that the scaling functions coincide favorably with those derived from the phenomenological model, Eqs. (\ref{eq:Mtilde})--(\ref{eq:chittilde}), which are highlighted by the thick black curves in Fig. \ref{fig:data} (e--g), even though we did not assume any specific form of scaling functions in the data collapse fitting procedure. The exponent $\alpha$ in the phenomenological scaling functions is obtained from the aforementioned data collapse fitting procedure, thus there are no more free parameters except for the overall proportional constant $c$. This remarkable consistency of scaling functions demonstrates that the low-energy physics of 2D RS state in the random $J$-$Q$ model is captured by the phenomenological model of random spin pairs. The one-parameter scaling and data collapse behavior of these thermodynamic quantities, and particularly the universal scaling functions can be taken as fingerprints of 2D RS states in experiments on quantum magnetic materials.

{\bf Summary and discussions.} The 2D RS state of the random $J$-$Q$ model in magnetic field is studied with large-scale QMC simulations. The magnetization and the susceptibilities show power-law dependence on magnetic field in the low-temperature limit. Moreover, these thermodynamic quantities exhibit one-parameter scaling and data collapse behavior with the ratio $B/T$, and the scaling functions are remarkably consistent with the phenomenological model of random spin pairs with a singular distribution of effective exchange interactions.

The cusp in $\chi_{L}$ shown in Fig. \ref{fig:data} (d) may reflect a quantum phase transition induced by the magnetic field. While $m_{z}(B,T)=-N^{-1}\pd_{B}F(B,T)$ is the first-order derivative of the free energy, the susceptibilities are given by $\chi_{L}(B,T)=\pd_{B} m_{z}(B,T)$ and $\chi_{T}(B,T)=m_{z}(B,T)/B$. Thereby, $\chi_{L}$ is a second-order derivative while $\chi_{T}$ is a first-order derivative of the free energy. Therefore, the cusp in $\chi_{L}$ may reflect a third-order quantum phase transition induced by the magnetic field, which cannot be clearly resolved in $m_{z}$ and $\chi_{T}$. The nature of the possible phase transition and the phase in the high magnetic field regime is left for future work.

The phenomenological scenario is borrowed from the asymptotically exact RSRG analysis of 1D RS states, which shows that quenched spatial randomness dominates and suppresses quantum fluctuations in the low-energy limit, resulting into the RS ground state with an essentially static configuration of spin singlet pairs instead of resonating valence bond states, i.e., coherent superposition of different spin singlet configurations. While the RSRG approach to 2D random spin systems is not as successful as in 1D \cite{Igloi2005} thus 2D RS states remain largely elusive, the remarkable consistency of the phenomenological model with our unbiased numerical results provides valuable guidance for more complete understanding of 2D RS states in the future. The dynamical critical exponent $z$ of the 2D RS is accessed with the finite-size scaling of the spin gap in a separate work \cite{Peng2024a}.

The one-parameter scaling and data collapse behavior of thermodynamic quantities has been observed in various QSL candidate materials and disordered antiferromagnets in recent years \cite{Kitagawa2018, Kimchi2018, Bahrami2019, Kundu2020, Do2020, Murayama2020, Song2021, Nguyen2021, Khatua2022, Murayama2022, Saha2023, Lee2023, Kang2023, Shivaram2024, Hossain2024, Lee2024, Bandyopadhyay2024, Bandyopadhyay2024a, Sana2024, Imamura2024, Guo2023}. In Ref. \cite{Li2019d}, the susceptibility of YbMgGaO$_{4}$ in magnetic field is found to reach a plateau at low temperature, and the plateau is higher at lower magnetic field. This feature is captured by the phenomenological model [see Fig. \ref{fig:data} (h)], because the divergent low-energy density of states is suppressed by magnetic field. We propose that a closer examination of the one-parameter scaling behavior and comparison with the phenomenological scaling functions can diagnose possible RS states in future experiments.

While we have focused on the 2D RS state with the full SU(2) spin rotation symmetry in the exchange interactions, the asymmetry and anisotropy in the magnetic exchange and the Zeeman interactions can modify the exponents in the scaling form \cite{Kimchi2018} and the universal scaling functions, which is left for future study.

\begin{acknowledgments}
This work is supported by the National Natural Science Foundation of China (Grant Nos. 12174387 and 12304182), Chinese Academy of Sciences (Nos. YSBR-057, JZHKYPT-2021-08, and XDB28000000), and the Innovative Program for Quantum Science and Technology (No. 2021ZD0302600).
\end{acknowledgments}

\bibliography{2DRS}
\end{document}